# Questioning Your Brilliance in Physics: Differential Shifts in Fixed Mindsets by Grade and Gender

Fargol Seifollahi[1], Christian D. Schunn[2], Chandralekha Singh[1]

*[1] Department of Physics and Astronomy, University of Pittsburgh, Pittsburgh PA 15260, USA*
*[2] Learning Research and Development Center, University of Pittsburgh, Pittsburgh PA 15260, USA*

**ABSTRACT**

Students' domain-specific mindsets and their beliefs about their capacity to improve through effort play a crucial role in shaping their experiences and decisions to persist in STEM disciplines. Physics is generally seen as a field requiring innate brilliance, which can reinforce fixed mindsets, particularly after initial setbacks in performance that are common in introductory university courses. In this study, we examine changes in fixed mindsets and potential gender differences in an introductory calculus-based physics course. Our sample consisted of 508 students with an average age of 18, predominantly White, with men comprising the majority. Based upon survey response distributions, three distinct mindset categories were identified: Hesitant, Hopeful, and Confident, describing how strongly students rejected a fixed mindset in physics. The results suggested large gender differences in distributions at the high and low-end groups. We also found an overall decline toward fixed mindsets across the course, and logistic regressions controlling for initial mindsets showed that women were significantly more likely than men to shift away from the Confident category. While the majority of men tended to stay within the Confident category, the majority of women moved away from it. Particularly, this differential shift was seen among students receiving Bs or Cs, the most commonly awarded grades in this course. Furthermore, there were relatively small differences in the probability of change within men as a function of grades received, whereas women showed marked declines toward fixed beliefs with either a B or C. Our findings provide empirical evidence for the dynamic, grade-sensitive nature of students' mindsets in a calculus-based physics course. In particular, gendered differences in the probability of retaining confidence in the face of commonly awarded lower grades have implications for improving instructional strategies and highlight the need for targeted interventions that help promote resilience among students at higher risks of adopting a fixed mindset.

## 1. Introduction

Much of the ongoing efforts in science, technology, engineering, and mathematics (STEM) education research have been focused on understanding students' experiences in foundational courses, where high attrition rates and representation gaps persist (Chen, 2013). Despite broad efforts to expand the pool of promising STEM talent, physics remains a field in which the growth in the attainment of bachelor's degrees has been comparatively slow relative to other STEM disciplines (Mulvey and Nicholson, 2020; Bowman et al., 2022). Physics also continues to be among the fields with the lowest percentage of bachelor's degrees awarded to women (National Science Board, National Science Foundation, 2021; Boatman, 2022), which has been associated with its reputation as a "brilliance-required" field (Scherr et al., 2017; Leslie et al., 2015). This under-representation persists in other closely aligned engineering disciplines, such as electrical or mechanical engineering (American Society for Engineering Education, 2023). These trends have been followed by a growing body of research examining psychological constructs that shape students' general academic performance and reactions to challenging STEM learning environments, one of which is intelligence mindset (Dweck, 2006; Williams et al., 2021).

The intelligence mindset theory conceptualizes people's views on the nature of intelligence: a fixed mindset and a growth mindset (Dweck and Leggett, 1988). Those with a fixed mindset think of intelligence as an immutable quality, whereas those with a growth mindset think of intelligence as something that can be cultivated through efforts, strategies, and the help of others. More recent work has emphasized domain-specific mindsets (e.g., immutable abilities in science and mathematics), showing that context-specific mindsets not only contribute unique predictive value to psychological and academic outcomes, but they also yield stronger associations with these outcomes than general mindset measures (Limeri et al., 2023; Limeri, 2025).

In the context of college introductory physics courses in the US, recent research has found pronounced gender differences in terms of students viewing their intelligence in physics as an innate quality, which appear to grow during these introductory courses (Malespina et al., 2022). These differences are likely a key part of lower persistence in STEM disciplines (van Aalderen-Smeets et al., 2019; Seymour et al., 2019). Additionally, studies that investigate similar yet distinct measures such as self-efficacy— confidence in one's *current* ability to succeed in a task—show that gender differences in self-judgments of competence in physics are much larger than the small differences in actual performance (Whitcomb et al., 2020).

It is important to note that self-efficacy remains conceptually distinct from mindset, even though these constructs are correlated (Wasylkiw et al., 2020). Self-efficacy specifically refers to confidence in one's current ability to perform specific tasks in a given domain (Bandura, 1997), whereas mindset concerns beliefs about whether one's abilities can fundamentally change through effort and persistence (Dweck, 2006). Although both constructs contribute to students' academic experiences, they capture different dimensions of self-perception; self-efficacy reflects judgments of present competence, while mindset concerns perceived potential for future growth. This distinction is critical for the present study; while previous findings have shown that men report significantly higher self-efficacy despite women often earning equal or better grades across most foundational STEM courses (Whitcomb et al., 2020), gender differences in changing mindsets and their relationship to grades remain largely unexplored.

The initial differences in mindsets could stem from earlier negative experiences inside and outside school classroom context (Steele, 1997; Beasley and Fischer, 2012). Within university contexts, it is necessary for classroom environments to be designed in ways that counteract these differential prior experiences rather than exacerbate them (e.g., shaping how students react to lower initial grades). Therefore, it is important to examine potential psychological constructs and underlying mechanisms, such as domain-specific mindsets, that may influence individuals prone to academic self-doubt at different stages of their academic journey.

Some previous research has focused on the role of domain-specific mindset in STEM disciplines, including some limited work on physics-specific mindsets (Dai and Cromley, 2014; Shively and Ryan, 2013). For example, beginning with a fixed mindset in physics predicted receiving lower grades in an introductory course, even controlling for prior academic performance (Malespina et al., 2022). However, previous studies have tended to examine changes in discipline-specific mindsets in terms of average declines towards fixed beliefs, and no study has explored shifts associated with grade feedback explicitly. These remaining questions make discipline-specific mindsets a relatively new area of investigation.

In the current research, we aim to understand not only how students' physics-specific mindsets change after taking an introductory physics course, but also explore how the direction and magnitude of this change may be different for men and women depending on their initial views, and whether similar grades convey the same information for both women and men in terms of predicting their changing mindsets.

## 2. Theoretical Framework

### *2.1. Intelligence Mindset*

The intelligence mindset theory, conceptualized by Dweck and colleagues, initially described beliefs about one's own abilities as existing along a spectrum with two contrasting perspectives (Dweck and Leggett, 1988; Dweck, 2006). At one end, a fixed mindset (also known as the entity theory of intelligence) corresponds to the belief that intelligence is fixed and not malleable. On the other end is the growth mindset (also called an incremental theory of intelligence), which sees intelligence as a malleable quality, and something that can be developed further through effort and persistence. Those with a fixed mindset were found to mostly rely on external feedback, judge their own ability relative to others, see mastery as indicating ability, and see effort as inversely related to ability (Nicholls, 1984; Muenks and Miele, 2017). Although intelligence mindset is correlated with measures such as self-efficacy or one's confidence about their current ability in a specific task or domain, mindset is fundamentally concerned with the potential for change in those abilities (Wasylkiw et al., 2020; Bandura, 1997), and has been linked to academic performance. Within introductory courses, this distinction between beliefs about current abilities (self-efficacy) and future possible abilities (mindset) is particularly important.

Given that students' attributions of struggle and their learning strategies are closely tied to their mindsets (Little et al., 2019), several interventions have sought to target students' mindsets, either implicitly or explicitly, to help students adopt more effective learning approaches (Yeager et al., 2019; Beatty et al., 2020; Samuel and Warner, 2021; Kramer et al., 2023; Yeager et al., 2022). A common theme across these interventions is helping students acknowledge adversity and struggle as a natural part of learning, and that striving against challenging problems is a means for expanding their knowledge and problem-solving skills (Burnette et al., 2023). One such intervention in college biology and physics classes successfully showed an increase in attendance rate, course grades, and 1-year college persistence, particularly for previously under-performing students (Binning et al., 2020). Mentoring students to view intelligence as malleable and exposing them to hardworking role models has also been effective in eliminating gender gaps in mathematics performance and an increase in standardized test scores for female students (Bagès et al., 2016; Good et al., 2003).

Such interventions are often brief and flexible in time, ranging from online modules to single classroom sessions. However, the degree of their effectiveness on academic achievement has been a topic of debate (Denworth, 2019; Yeager and Dweck, 2020). Recent meta-analyses on mindset interventions found inconsistent effect sizes across studies, with the majority of them indicating small to null effects (Sisk et al., 2018). At the same time, the same meta-analyses found evidence supporting the effectiveness of mindset interventions for academically high-risk students and those coming from a low socioeconomic status. It is therefore crucial to understand the heterogeneity of effects, i.e., for whom and when such growth mindset interventions work, as intervention effects may differ in initial ideally-controlled versus later broader scaling attempts (Greenberg and Abenavoli, 2017).

It is also important to recognize that fostering the idea of malleable intelligence is more complex than simply introducing students to the growth mindset. It may be possible that students hear about the neuroplasticity of the human brain and accept the idea that intelligence can be cultivated in a general sense. However, they may still remain unconvinced about their own capacity to do so on a personal level. This distinction between the concept of self versus others' capacities has been highlighted in a previous research work, where researchers proposed a revised "self-theory" measure of intelligence. They found that students' personal beliefs about their own ability to develop intelligence were stronger predictors of academic outcomes than their general beliefs about the malleability of intelligence (De Castella and Byrne, 2015). Such findings bring attention to the importance of distinguishing between beliefs about self and broader views on intelligence, especially when analyzing survey data or evaluating the effectiveness of interventions.

Another essential point to consider is how mindsets can vary across different academic disciplines. Physics, in particular, is a field that has been associated with brilliance stereotypes (Scherr et al., 2017), and the under-representation of women in the field has led to a scarcity of diverse role models for later generations pursuing physics (Cheryan et al., 2017). Such societal narratives can decrease students' self-efficacy, i.e., their confidence in their current ability to succeed, and may also further perpetuate fixed mindsets or beliefs about the malleability of abilities (Bandura, 2012). As noted earlier, although self-efficacy and mindset are conceptually distinct, they are often correlated and can be shaped by similar environmental cues (Wasylkiw et al., 2020). Specifically in introductory physics courses, previous research has shown that women's performance is positively predicted by their levels of self-efficacy above and beyond prior preparation (Sawtelle et al., 2012; Malespina et al., 2024a), and that at the end of calculus-based introductory physics courses women with A grades have similar physics self-efficacy as men who get a C grade (Marshman et al., 2018). While these findings underscore the importance of self-efficacy as a reflection of students' perceived current competence and its relation to performance, they also raise a related but distinct question of whether similar disciplinary environments and performance signals shape students' mindsets, that is, their beliefs about their capacity for future growth.

*2.2. Physics Intelligence Mindset*

Intelligence mindset theory has been a subject of interest in educational research. Recent work extending mindset theory across STEM disciplines has demonstrated that science and math-specific mindset measures predict psychological and academic outcomes more strongly than general mindset measures (Limeri, 2025). For instance, domain-specific studies in the mathematics context have shown that students' math mindsets are associated with motivation and achievement, and growth beliefs support better grades and motivation (Cribbs et al., 2021; Blackwell et al., 2007). Furthermore, longitudinal research in math and computer science suggests that students' domain-specific mindsets can become more fixed over a semester and that more fixed beliefs are associated with lower performance and engagement (Shively and Ryan, 2013; Flanigan et al., 2017).

Still, fewer research efforts have focused on physics-specific mindsets (Goldhorn et al., 2023; Spatz and Goldhorn, 2021). In a study exploring this gap, researchers conducted interviews with students enrolled in introductory physics courses with the goal of understanding how students identify challenges in learning and how their response to challenges aligns with the mindset literature (Little et al., 2019). Emerging evidence suggests that mindset views go beyond a simple one-dimensional spectrum with two opposite extremes (i.e., fixed vs. growth). The same interview-based study uncovered instances in which students endorsed both fixed and growth mindsets. For example, one student mentioned that they were "bad at physics" because it does not simply come as easy to them as their peers, reflecting a fixed mindset with an emphasis on innate ability. However, the same student acknowledged later in the same interview that hard work and seeking help during office hours had improved their understanding of physics. Similarly, interviews with physics faculty revealed the simultaneous existence of both views about physics students (Scherr et al., 2017).

As noted earlier, this duality is not limited to self-perceptions but can manifest as students having different views about their capacity to develop their own intelligence as opposed to their peers' capacity. In view of findings that challenge the one-factor continuum model of mindset, several studies have explored the nature of mindset as a multidimensional scale (Ortiz Alvarado et al., 2024; Troche and Kunz, 2020). More specifically, recent research in the context of introductory calculus-based physics courses revealed four distinct mindset categories through multidimensional scaling analysis (Kalender et al., 2022). This study examined two dimensions, effort versus ability and self versus others, and identified four separate mindset views, which in this paper we refer to as "my fixed beliefs", "my growth beliefs", "others' fixed beliefs", and "others' growth beliefs". The resulting four-factor model was also re-validated in other studies in the context of introductory physics courses (Malespina et al., 2022, 2023). Results from these studies indicated that the "my fixed beliefs" factor, i.e., the extent to which students accept or deny a fixed mindset about themselves, was the strongest predictor of course grades and also where the largest gender differences were observed.

*2.3. Reactions to Grades Feedback*

A large topic in motivational research involves how students respond to performance feedback (Pitt and Norton, 2017; Ryan and Henderson, 2018; Canning et al., 2024). Within research on intelligence mindsets, overall and in physics, the focus has been on mindsets as predictors of future behaviors (Haimovitz et al., 2011) and performance (Costa and Faria, 2018). Such outcomes are not only predicted by students' own beliefs about the nature of intelligence, but can also be shaped by faculty mindsets and students' perceptions of their beliefs, especially among underrepresented students (Canning et al., 2019; Rattan et al., 2018).

These mindsets can change, particularly in response to different learning experiences and types of feedback provided (Dai and Cromley, 2014; Limeri et al., 2020). For example, instructional practices such as providing formative assessments and opportunities to revise work may serve as cues that the instructors believe in students' potential for growth. On the other hand, if students rely only on a few heavily weighted exams, it is possible that they will interpret lower grades as a sign of incompetence. Such a reaction would be more pronounced if students receive lower grades than they are used to, e.g., introductory university course grades in their first year, in contrast to the often higher grades they had in high school (Westrick et al., 2023).

It is also likely that student mindsets will change as they try to make sense of why their performance has declined despite their problem-solving and studying behaviors being the same as in high school coursework. However, this has not been directly studied. It may be that some students attribute poor performance to the changing requirements and necessary learning strategies (i.e., by maintaining or even developing a growth mindset), whereas other students attribute poor performance to a changing understanding of how fundamentally difficult physics is (i.e., by developing a fixed mindset).

Prior work on examining the more fine-grained structure of motivational changes with experience shows that small changes in average performance in a population can mask different patterns of change found within a subset of students (Wang et al., 2017). Investigations of individual differences in change are particularly relevant to studying changes by particular populations, such as women in physics. For a number of reasons, women might be more likely to show substantial change in fixed mindsets. First, they are more likely to get lower grades on exams, considering different experiences of academic and social challenges (Matz et al., 2017; Salehi et al., 2019). Second, microaggressions by fellow students or the experience of being a numerical minority in the physics classroom may lead them to interpret poor exam performance as a lack of potential for success (Barthelemy et al., 2015; Scherr et al., 2023). Further, this gendered reaction to poor outcomes might only matter in especially poor performance situations (e.g., receiving a C or D grade).

In this work, we use previously validated survey questions to study not only the progression of students' mindsets about their own physics abilities, but also the direction and degree of these changes and the role that different levels of grade feedback might play in shifting or reinforcing initial mindsets. While much of the existing literature has examined how mindset predicts course outcomes and gendered differences in domain-specific mindset (Limeri et al., 2023; Heyman et al., 2002), how students' perceptions of course grades differ in terms of predicting changes in their mindset in introductory physics remains unexplored. Below, we provide our research questions and directional hypotheses for identifying at-risk students:

RQ1. Does the distribution of students' mindsets about their own abilities in physics change over the course of taking an introductory physics course, and are there any gender differences in the distributions at each time point?

H1a. Students are expected to develop more fixed mindsets regarding their physics abilities,

H1b. Women may be more likely to develop these fixed mindsets.

*Reasoning:* In the absence of explicit interventions promoting a growth mindset, a negative shift towards a fixed mindset is often observed in STEM courses (Limeri et al., 2020; Yeager et al., 2019; Shively and Ryan, 2013). Additionally, as prior research shows that women in STEM contexts are more likely to endorse fixed beliefs about their abilities, we therefore expect a gender difference in the introductory physics context (Heyman et al., 2002).

RQ2. What patterns of physics mindset shifts occur, and how do these patterns differ between men and women?

H2a. Patterns of change will include a mixture of growth, small declines, and large declines,

H2b. Large declines will be more common in women.

*Reasoning:* The size and direction of change will depend upon student experiences in the class. When women experience major challenges, they may be more likely to interpret those challenges as internally caused based upon stereotypes (Beasley and Fischer, 2012).

RQ3. Does grade predict different patterns of change in physics mindset, and do these relationships between grades and mindset change differ between men and women?

H3a. Students with lower grades will gravitate more towards fixed beliefs,

H3b. Women will be more likely to show this transition after lower grades.

*Reasoning:* Grades are one salient source of performance feedback, and lower performance (grades or others) after significant effort can be taken as evidence of fixed abilities (Limeri et al., 2020). Evidence in the engineering context suggests a greater sensitivity to even moderately poor performance among women (Heyman et al., 2002).

To answer our research questions, we focus on the "my fixed beliefs" factor from the previously described four-factor framework over the course of a two-semester sequence of calculus-based introductory physics courses for physical science and engineering students. Details are further elaborated in the following section.

## 3. Methodology

### *3.1. Participants and Course Structure*

This study was conducted at a large research university in the United States. The participants were physical science and engineering students who are typically required to take introductory calculus-based Physics 1 and 2 as part of their academic program. The Physics 1 course primarily covers the following topics: kinematics, forces, work and energy, rotational kinematics and dynamics, simple harmonic motion, gravitation, and waves. The topics primarily covered in the Physics 2 course are: electrostatics, magnetostatics, behavior of simple electrical circuits, and wave optics.

Data were collected for two consecutive academic years (2022-2023 and 2023-2024) for students who took Physics 1 in the Fall and Physics 2 in the Spring semester. Students across these cohorts had the same set of instructors for the Physics 1 course, the same course content, and a shared standardized final exam at the end of the course. Students taking Physics 1 and Physics 2 in the summer or the opposite semester (i.e., taking Physics 1 in the Spring semester or Physics 2 in the Fall semester) were not included in the analysis because there were much smaller numbers of those students and they likely had substantially different patterns of motivation and motivational change. For example, a majority of the students taking Physics 1 in the Spring semester or Physics 2 in the Fall semester are ones that are repeating the course. Also, many students enrolled in the summer courses are students attending other institutions during the school year.

The Physics 1 course shared a common structure across semesters and instructors in terms of assessment. A larger portion (approximately 60%) of the overall course grade was determined by performance on three midterm exams (which included feedback comments) and one common cumulative final exam across all instructors. Throughout the semester, students were given weekly multiple-choice quizzes that they could complete collaboratively. Students received grade feedback on quizzes, and teaching assistants would often review any quiz problems that the majority of students answered incorrectly.

For our analysis, we focused on students who had successfully passed Physics 1 and took Physics 2 in the following semester since those students completed the survey after receiving grades from Physics 1. This left us with $N = 626$ students who had completed at least a pre-survey (beginning of Physics 1) or post-survey (beginning of Physics 2), and had successfully passed an attention check item which asked them to choose option "Disagree" for that item. Among these students, 94% of them took the pre-survey and 87% of them took the post-survey. Excluding the 36 and 82 students missing responses for the mindset items of interest to us on the pre and post surveys, respectively, we ended up with a sample size of $N$ =508 students for analysis purposes.

86% of students were enrolled in the School of Engineering, and they were almost entirely first-time freshmen. In terms of racial composition, the sample consisted of 66% White, 22% Asian, 9% Black/Latiné/Indigenous, and 3% unknown, according to university data. Women comprised 39% of the sample compared to 61% men, which are typical percentages for gender distribution in introductory calculus-based physics courses at this institution in recent data.

### *3.2. Measures and Procedures*

#### *3.2.1. Physics Mindset*

Attitudinal surveys were administered during the first and the last week of classes. The surveys were available online during those weeks, and students were reminded to take them during the recitations. Critically, this means the students had not yet received the final grades when they completed the post survey given within the same class (i.e., Physics 1). Therefore, we focused on the survey collected at the beginning of Physics 1 as the pre-survey and the survey collected at the beginning of Physics 2 as the post-survey. We chose not to contrast the end of the first semester and with the beginning of the second semester, because, although final grades are provided during that time, at the end of the first semester, much of the grade feedback had already occurred before the final exam via midterms and quizzes.

Physics mindset items were adopted from a survey previously validated for introductory calculus-based physics students (Malespina et al., 2022). The larger survey included four distinct and relatively independent views on mindsets in physics that were revealed through multidimensional scaling analysis (Kalender et al., 2022). All of the prior studies that serve as the foundation for our work found that the factor focused on one's fixed beliefs about themselves had the largest gender differences at the beginning of university instruction and was also the best predictor of course grades. Therefore, for this work, we focused on "my fixed beliefs" construct: students' tendency to either reject or endorse a fixed mindset about their own physics abilities.

In all survey items, "physics" was explicitly mentioned to ensure that the survey questions were measuring the domain-specific mindset. The four items are shown in Table 1. All items were on a seven-point Likert scale (Strongly Disagree = 1, Disagree = 2, Slightly Disagree = 3, Neutral = 4, Slightly Agree = 5, Agree = 6, Strongly Agree = 7) and were then reverse-coded, so that a higher score indicates the desired attitude of rejecting fixed mindset in physics. Cronbach's alpha for this measure was 0.91, indicating high reliability.

To enhance the precision of our interpretations and ensure a more accurate representation of the latent variable of interest, we first calculated factor scores for "my fixed beliefs" construct using the "lavPredict()" function from package "lavaan" in R Studio version 2024.04.2+764 (Rosseel, 2012; R Core Team, 2024). Using these factor scores accounts for the complex relationships among the items within that factor and incorporates item loadings into the predicted construct score rather than treating all items equally as in a simple average score.

Table 1: Mindset items focused on “my fixed beliefs” included in the survey. The same survey items were given to the students in both courses. All items were reverse coded for analysis.

| **My Fixed Beliefs** ($\alpha = 0.91$) |
|---|
| 1. Even if I were to spend a lot of time working on difficult physics problems, I cannot develop my intelligence in physics further. |
| 2. I won’t get better at physics if I try harder. |
| 3. I could never excel in physics because I do not have what it takes to be a physics person. |
| 4. I could never become really good at physics even if I were to work hard because I don’t have natural ability. |

The factor scores showed a negative skewness and multimodality that was also observed in the majority of the raw items. A number of methodology texts recommend treating data with multiple modes as categorical data representing distinct subgroups (Dodge, 2008; Bulmer, 2024). Based on the observed distribution patterns, we categorized students into three mindset groups using cutoff values that were derived from local minima between modes in the response frequency distribution. Given our relatively large sample size, we were able to use Kernel Density Estimate (KDE) plots in R to visualize the response frequency distribution and determine local minima (Bramson et al., 2016; Węglarczyk, 2018; Kim, 2021; Silverman, 2018). Based upon the derived cutoff values, we labeled the three groups as Hesitant, Hopeful, and Confident. Within this population of students and given the negative skewness of the response frequency distribution, no group could be meaningfully labeled as strongly endorsing a fixed mindset, whereas many students strongly rejected the fixed mindset or were unsure. Consistent with their labels, the average raw scores for the Hesitant, Hopeful, and Confident categories were 4.5, 5.8, and 6.7, respectively, on the 1-7 point Likert scale. Other approaches to category definitions are also possible; for example, using equal proportions (i.e., quantiles), which is more common in educational research settings (Gelman and Park, 2009), or fixed Likert scale thresholds. Although providing a weaker fit to the data, these alternative approaches produced similar conclusions. Implementation details and findings using these alternative approaches are provided in the supplementary material.

*3.2.2. Demographic Information*

Demographic data were obtained from university records, which were based on the information students provided at the time of their enrollment in the university. Students could declare their race/ethnicity using the following options: White, Asian, Black, Hispanic, American Indian, or Pacific Islander. Multiple selections were allowed for students identifying as multiracial. For reporting purposes, we combined the race/ethnicity groups that each comprised about or less than 5% of the sample into a single category. Nine students declared a gender that did not match their sex at birth. These individuals were not included in our study due to an insufficient number for quantitative analysis.

*3.2.3. Course Grades*

Course grades at this university were letter grades (A-F), with + and - possible for each letter except an F. For analysis and to have enough cases for each specific grade, course grades were binned into A, B, or C grades, where + and - were ignored (e.g., the A bin includes A+, A, and A- grades). Since we matched students’ pre (beginning of Physics 1) and post scores (beginning of Physics 2), all of the students in our sample had obtained at least a C grade in Physics 1 in order to advance to Physics 2.

### 3.3. Analysis

The student identifiers in the collected survey data were converted into researcher IDs by an honest broker in order to be integrated into an analysis file with the similarly indexed demographic information provided by the university. Given the categorical nature of the fixed mindsets, the analysis of changing mindsets in response to receiving different grades involved logistic regressions (Theobald et al., 2019). In particular, to explore transition patterns by gender across different mindset categories and their significance level, logistic regressions were used to predict, by gender, the odds of students transitioning into each of the three mindset categories by the beginning of Physics 2, given a certain initial mindset category at the beginning of Physics 1. We similarly conducted logistic regressions that added Physics 1 grade and the interaction of gender and grade. Dummy variables for each mindset category were created using the package "fastdummies" in R (Kaplan and Schlegel, 2023; R Core Team, 2024). The logistic regression models were implemented using Stata v18 (StataCorp., 2023).

Additionally, since students were nested within four classes with different instructors, we examined possible effects of classroom-level nesting by testing hierarchical models that included the instructor variable both as random and fixed effects. We found the intra-class correlation coefficient (ICC) to be very close to zero, consistently explaining less than 0.3% of the variance in outcomes, and the fixed effects of the instructor variable remained insignificant. Therefore, the non-nested models were used for the primary analyses.

We also examined whether survey data were missing at random or if there was a systematic pattern of missingness. We found a marginal correlation between students not completing the survey at the beginning of Physics 1 and their final grade in the course (additional detail provided in supplementary material), both with and without controlling for additional predictors. Therefore, to ensure the robustness of our results, analysis was carried out first using only complete cases and then again using all students with data imputation for missing values. Missing values were imputed using multivariate imputation by chained equations with the MICE package in R (van Buuren and Groothuis-Oudshoorn, 2011; R Core Team, 2024). The findings showed similar patterns and were nearly identical; therefore, we report the simpler complete-case analyses without imputation. The patterns in distributions across grades and mindset categories also remained similar when we expanded the analysis to all available students for descriptive statistics. We also used students' advanced placement (AP) Physics scores as an additional variable that may explain gender differences observed in our analysis. However, the results remained similar, so we chose to present the analyses without AP Physics predictors, given the relatively high rates of missing AP data and the likely indirect effects of that experience already being captured by mindsets measured at the beginning of Physics 1.

### 3.4. Ethical Considerations

The study was conducted in accordance with the Declaration of Helsinki and approved by the University's Institutional Review Board. The research was approved as exempt, and the written informed consent requirement was waived. Students' participation in the attitudinal surveys was voluntary. Survey instruments were initially collected with student identification numbers to enable matching with university data. Then, all identifiers were anonymized by an honest broker prior to analysis to ensure confidentiality. None of the researchers involved in this study were instructors for the participating students, and all analyses were conducted on de-identified data.

## 4. Results

### 4.1. RQ1: Overall Distributions by Course and by Gender

To answer our first research question, we examined the descriptive statistics for our student sample. Figure 1 shows the sample sizes and the average raw scores from the surveys administered during the first week of classes for both Physics 1 and Physics 2, disaggregated by gender. The scores presented here are based on the original 1-7 point Likert scale before being transformed into factor scores. We used non-parametric tests to measure the effect size of differences and their significance. For comparisons between genders, this included the use of Cliff's Delta (Meissel and Yao, 2024) and rank-sum tests (Wilcoxon, 1992). For comparisons within gender groups, we used rank-biserial correlations (Cohen, 2013) and sign-rank tests (Wilcoxon, 1992). Overall, there were statistically-significant, small-to-medium-sized gender differences in fixed physics mindset scores at the beginning of both the Physics 1 and Physics 2 courses. Also, the means for both women and men showed small yet statistically-significant declines towards fixed beliefs going from Physics 1 to Physics 2. The average score comparisons between and within gender groups can be found in Figure 1.

| | Women | | | Men | | | Cliff's Delta |
|---|---|---|---|---|---|---|---|
| | N | Mean | SD | N | Mean | SD | |
| **Physics 1** | 199 | 5.7 | 1.1 | 309 | 6.1 | 0.8 | 0.21*** |
| **Physics 2** | 199 | 5.4 | 1.2 | 309 | 5.8 | 1.0 | 0.24*** |
| **Rank-Biserial $r$** | 0.27*** | | | 0.27*** | | | within group / between group |

Figure 1: Sample size ($N$), mean fixed physics mindset scores, and standard deviations (SD) for women and men at the beginning of Physics 1 and Physics 2. Comparison of average scores is reported both between gender groups in each course, as well as for each gender over time. ($^{***} = p < 0.001$)

Overall changes in mean might reflect changes in parts of the distribution (e.g., greater representation in the middle vs. the low part of the distribution). The percentages of men and women in each mindset category are shown in Figure 2. Here, there were large differences by gender in representation in the Hesitant group, with almost twice as many women represented in that group at the beginning of both Physics 1 and 2. Complementary and statistically significant percentage differences were observed at the high end, with men being in the Confident group approximately one-third more often than women. Thus, the medium-sized mean differences discussed earlier obscured larger differences in representation at the ends of the distributions. In addition, comparing the distributions across courses, roughly similar numbers of students are seen within the Hopeful category. The main distribution differences between courses are that, by the beginning of Physics 2, there are significantly more Hesitant students (a significant increase for both men and women) and fewer Confident students (a significant decrease for both men and women), as determined by McNemar Chi-Square statistical tests.

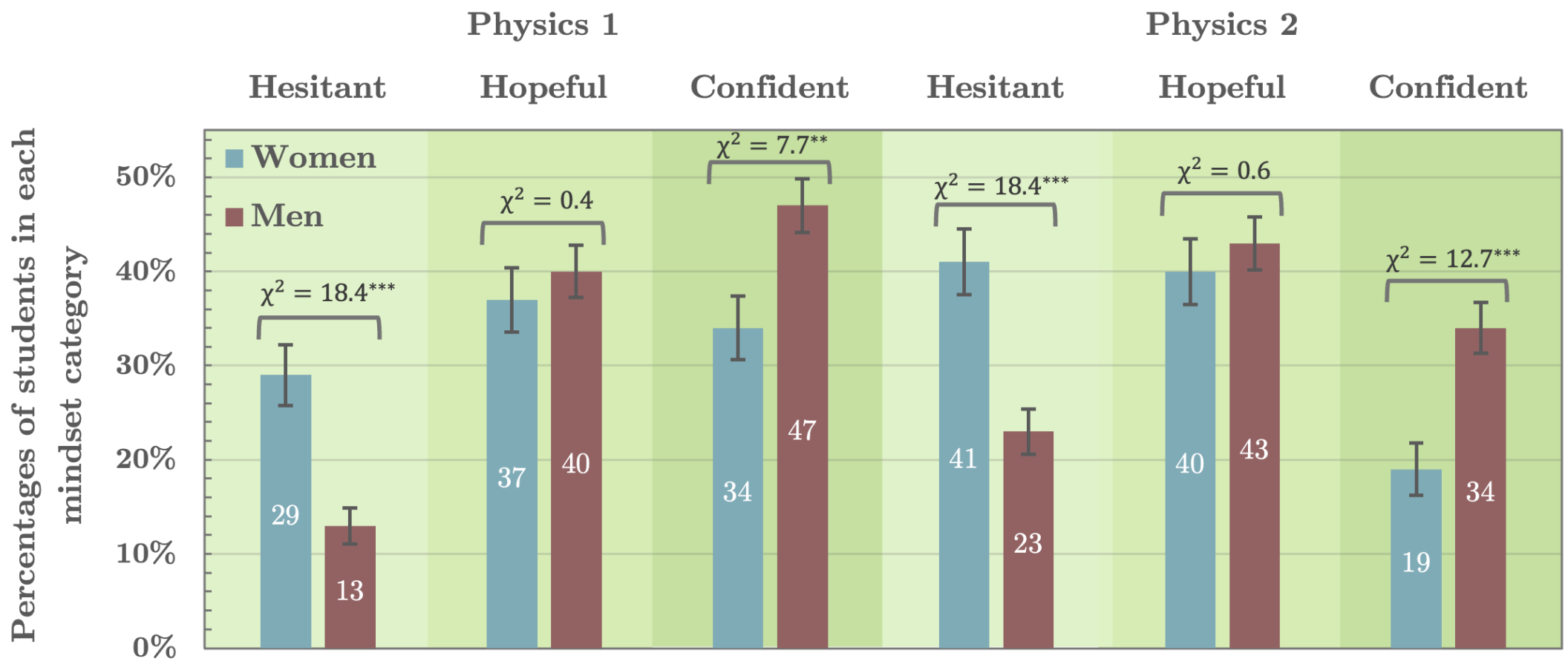

Figure 2: Percentages of men and women in each mindset category at the beginning of Physics 1 and Physics 2. Chi-Square statistics are reported as a measure of differences in representation by gender. ($^{**} = p < 0.01$, $^{***} = p < 0.001$)

*4.2. RQ2: Transition Patterns*

Changes in overall distributions of category membership at two time points can mask additional bidirectional changes that occur. Figure 3 illustrates the relative frequency of transitions between different mindset categories (i.e., Hesitant, Hopeful, and Confident), separately for men and women. The transitions are represented by nine arrows for each gender, representing each of the three possible originating (beginning of Physics 1) and three possible ending (beginning of Physics 2) categories. Percentages along the arrows denote the percentages within each starting point (i.e., arrows emanating from each starting point add to 100%). Overall, the most common pathways were staying within the same category. However, this was often barely a majority, and many students changed categories, especially moving from higher mindset categories to lower mindset categories.

Faded arrows represent non-significant gender differences in terms of the likelihood of making that particular mindset transition, based upon the logistic regression results presented in Figure 4. Two specific transition pathways showed statistically significant differences by gender, and both involved transitions from the Confident category. First, men were 16% more likely to remain in the Confident category than were women. Second, women who were initially in the Confident category were 12% more likely to transition all the way into the Hesitant category than were men. This was a very large effect, with women being more than twice as likely to show that transition.

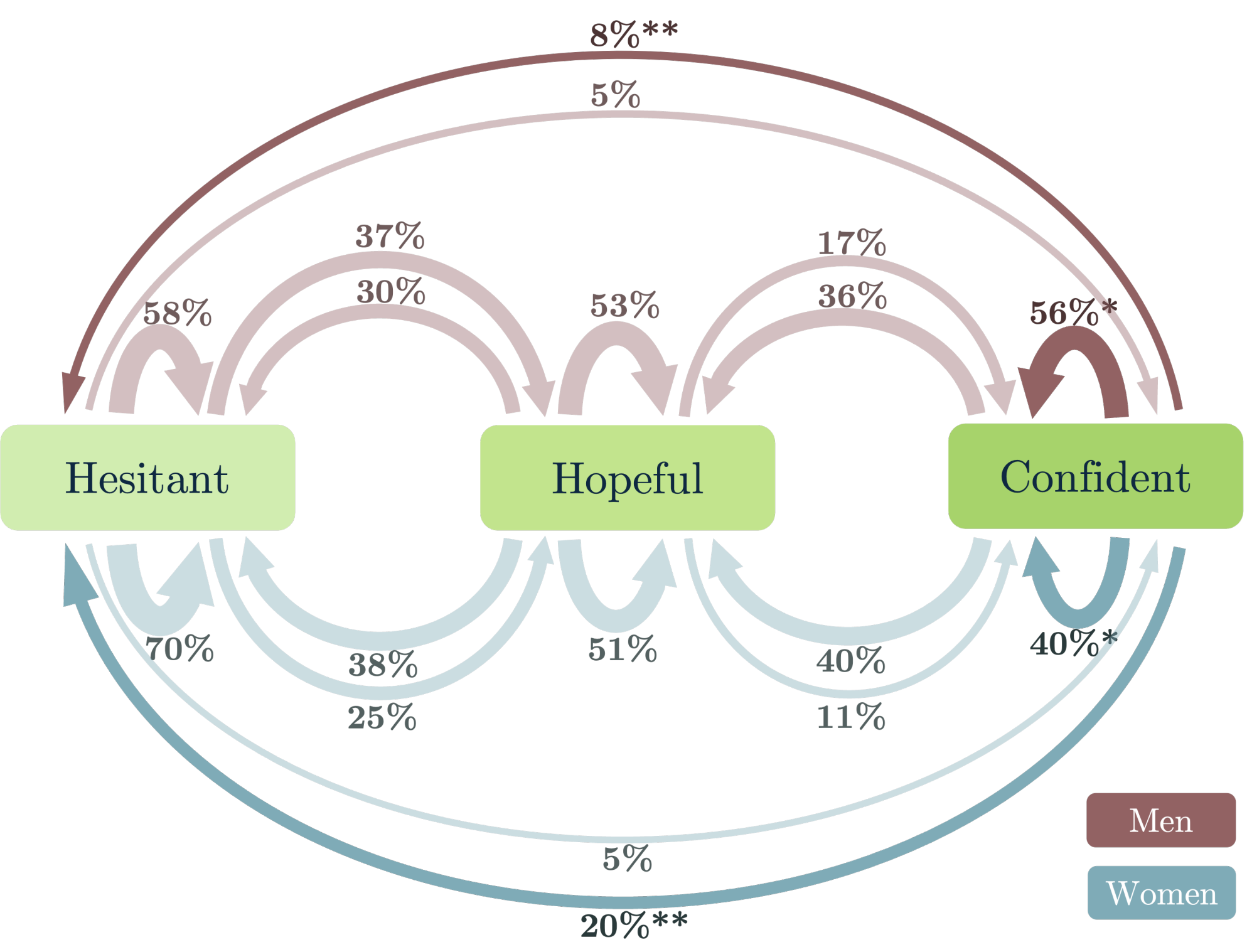

Figure 3: Transition patterns by gender. The arrows emanate from the mindset category at the beginning of Physics 1 and end at the mindset category at the beginning of Physics 2. (* = $p < 0.05$, ** = $p < 0.01$)

| Starting \ Ending | Hesitant | | | Hopeful | | | Confident | | |
|---|---|---|---|---|---|---|---|---|---|
| | OR | SE | $p$-value | OR | SE | $p$-value | OR | SE | $p$-value |
| **Hesitant** | 0.60 | 0.26 | 0.234 | 1.77 | 0.79 | 0.201 | 0.92 | 0.86 | 0.932 |
| **Hopeful** | 0.70 | 0.22 | 0.247 | 1.08 | 0.32 | 0.798 | 1.68 | 0.75 | 0.242 |
| **Confident** | 0.32 | 0.14 | 0.008 | 0.86 | 0.30 | 0.613 | 1.95 | 0.58 | 0.025 |

Figure 4: Odds ratios (OR), standard errors (SE), and $p$-values based on logistic regressions conducted to examine transitions across mindset categories. *Note:* For each case, the outcome is a binary indicator of the ending category, controlling for gender (reference group: women) and given a particular starting category. For example, the odds of men starting and remaining in the Confident category are 1.95 times higher than those of women (1.95:1)

### *4.3. RQ3: Reactions to Grades*

Addressing our third research question, we then examined what might underlie these gendered changes in fixed beliefs from Physics 1 to Physics 2. In particular, since women often obtain somewhat lower grades in Physics 1, such a difference in experience might have signaled to them that they had lower abilities than initially thought. Alternatively, since the grade differences were smaller than some of the large attitudinal differences, women might have reacted differently to grade information than did men. We first examine the grades they received and then turn to the potential reactions to those grades.

Figure 5 shows the distribution of grades for all students who took the Physics 1 course, as well as the distribution specifically for those who passed the course with a minimum grade of C. We note that since the minimum passing grade for the Physics 1 course was a C and a C- is considered a failing grade, those with a C- are included in the DFW (drop, fail, or withdraw) category rather than C. On the left side of Figure 5, percentages in each grade bin are shown for all students who took Physics 1 in a Fall semester, including those who dropped, failed or withdrew from the course. On the right side, we have represented the grade distribution for students from the sample with matched pre and post mindset survey responses who had successfully passed the Physics 1 course.

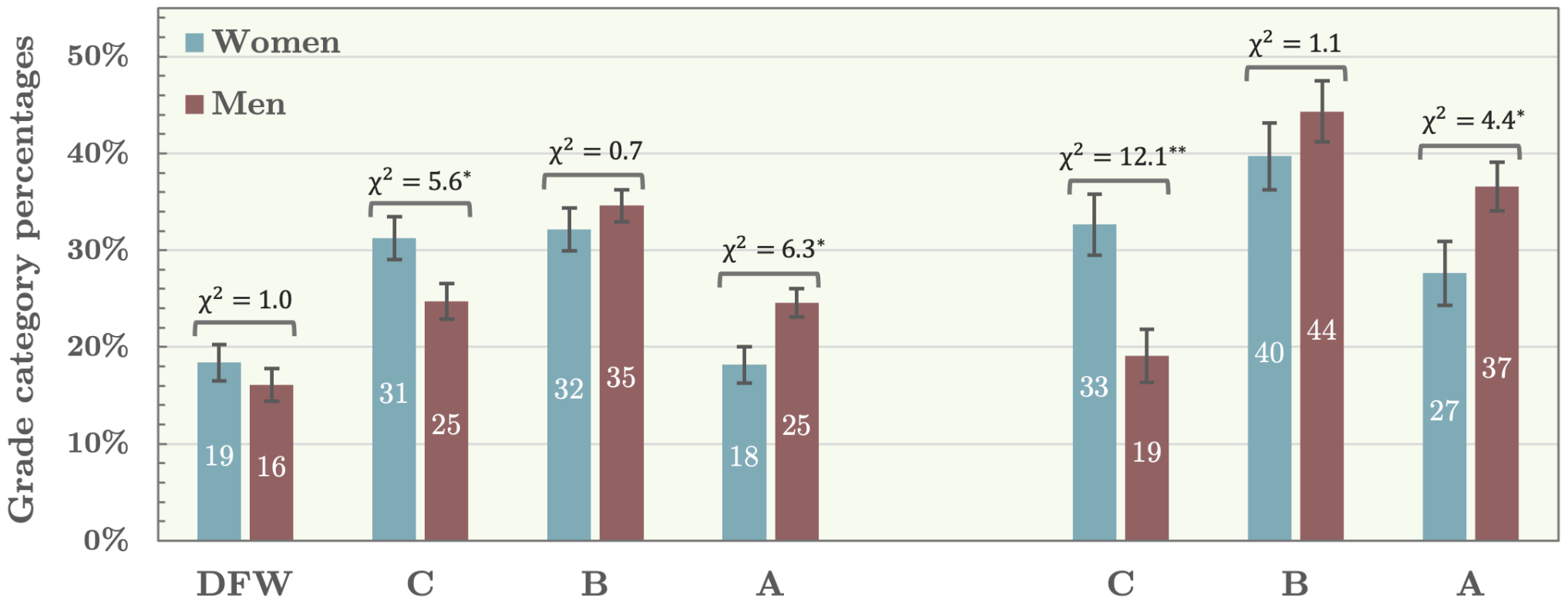

Figure 5: Grade distributions by gender for all Physics 1 students vs. students who advanced to Physics 2. *Note:* In the first set, $N$=441 for women and $N$=663 for men. For the sample with matched pre and post scores, $N$ = 199 for women and $N$ = 309 for men. Chi-Square statistics are reported as a measure of differences in representation by gender. ($* = p < 0.05$, $** = p < 0.01$)

Comparing percentages of men and women in each grade category for all students taking Physics 1, Chi-Square tests revealed significant gender differences in proportions within the A and C categories, but not in B or DFW categories. The results show that while there is no statistically significant difference between men and women in terms of advancing to the Physics 2 course, women tend to get more Cs and fewer As as compared to men, which represents the two extremes of passing grades. Past work suggests these small but reliable gender differences in performance vary by instructional format (Lorenzo et al., 2006), are not as pronounced in other introductory science courses (Taasoobshirazi and Carr, 2008), and may be explained by motivational rather than skill or knowledge differences. In the current study, we do not seek to explain the differences, but simply need to report the grade distributions to examine their potential influence on student mindsets: there were sufficient numbers of women and men students at each grade level to examine differential reactions to these grades.

In particular, we examine the results of the logistic regression models, where we added grades and the interaction of grades and gender as predictors in addition to initial mindset. Here, we only focused on the high and low ends of the mindset categories, i.e., Confident and Hesitant, given the large gender differences in either staying in the Confident group or transitioning away from it to the Hesitant group. The details of the models used can be found in Table 2. Pseudo-R-squared values for the models for Confident and Hesitant outcomes were 0.23 and 0.19, respectively, representing excellent and nearly excellent fits (McFadden, 2021).

At the top of the table are simple main effects. Not surprisingly, students who were initially categorized as Hesitant were less likely to confidently reject a fixed mindset in Physics 2, and those who were initially Confident were much more likely to remain Confident (see Table 2 for reference group information). Similarly, students initially in the Hesitant category were more likely to stay within that group and unlikely to transition to the Confident category.

Next, the table shows the overall effects of grades. While the directions of the effects are not surprising, their magnitudes show evidence of large effects. In particular, students who received a B or C were much less likely to be in the Confident category and much more likely to be in the Hesitant category. Note that this effect controls for the starting position, and thus represents changes in mindsets relative to obtained grades.

There were no overall statistically significant gender differences in the transitions into Confident or Hesitant. Thus, the observed large differences in these transitions were localized to having obtained specific grades. In particular, there was a large additional effect of being a man with a B or a C grade with odds ratios of approximately 3 and 5, respectively, significantly increasing the likelihood of being categorized as Confident compared to women with similar grades. Interestingly, there were no significant gender interactions with grade for ending in the Hesitant category. Instead, that status appeared to primarily reflect the initial mindset and grades.

Table 2: logistic regression models used for analysis, with binary outcomes of finishing in the Confident/Hesitant category at the beginning of Physics 2. Odds ratios (OR), standard errors (SE) and *p*-values are reported for the predictors, i.e., mindset at the beginning of Physics 1, grade, gender, and grade and gender interaction terms.

| Variable | | Confident | | | | Hesitant | | | |
|---|---|---|---|---|---|---|---|---|---|
| | | OR | SE | z | $p>\|z\|$ | OR | SE | z | $p>\|z\|$ |
| Prior mindset placement (reference category: Hopeful) | Hesitant | 0.39 | 0.20 | -1.84 | 0.066 | 3.27 | 0.88 | 4.39 | 0.000 |
| | Confident | 6.26 | 1.59 | 7.24 | 0.000 | 0.28 | 0.08 | -4.72 | 0.000 |
| Grade (reference grade: A) | B | 0.16 | 0.08 | -3.72 | 0.000 | 2.47 | 1.07 | 2.10 | 0.036 |
| | C | 0.10 | 0.06 | -3.96 | 0.000 | 3.73 | 1.67 | 2.94 | 0.003 |
| Gender (reference category: Women) | Men | 0.77 | 0.29 | -0.70 | 0.487 | 0.74 | 0.34 | -0.65 | 0.514 |
| Interactions | Men with B | 3.38 | 1.94 | 2.12 | 0.034 | 0.75 | 0.42 | -0.52 | 0.601 |
| | Men with C | 5.15 | 3.62 | 2.34 | 0.019 | 0.74 | 0.45 | -0.50 | 0.618 |
| Intercept | | 0.38 | 0.13 | -2.79 | 0.005 | 0.31 | 0.12 | -3.14 | 0.002 |

To show the combined effects of grade and gender on ending mindset, we present marginal means by gender and grade in Figure 6. In particular, we used the previously fitted logistic regression models to calculate marginal means at each level of the interaction between grade and gender. The results suggest that there are no significant gender differences in being classified as Confident in Physics 2 after receiving an A. Also, it is clear that the probability of being Confident decreases for both men and women as grades fall below an A. However, women show a much more pronounced reaction to receiving a B or C grade, with a statistically significantly larger decrease in their likelihood of being categorized as Confident compared to men in Physics 2. More specifically, women with a B grade are almost half as likely as men with a B grade to identify as Confident in Physics 2, while women with a C grade are nearly three times less likely than men with the same grade to fall into the Confident category. This suggests that grade feedback may have a more substantial influence on women's confidence in rejecting fixed beliefs compared to men, and highlights the significant role that academic performance might play in shaping physics mindset as students advance to Physics 2.

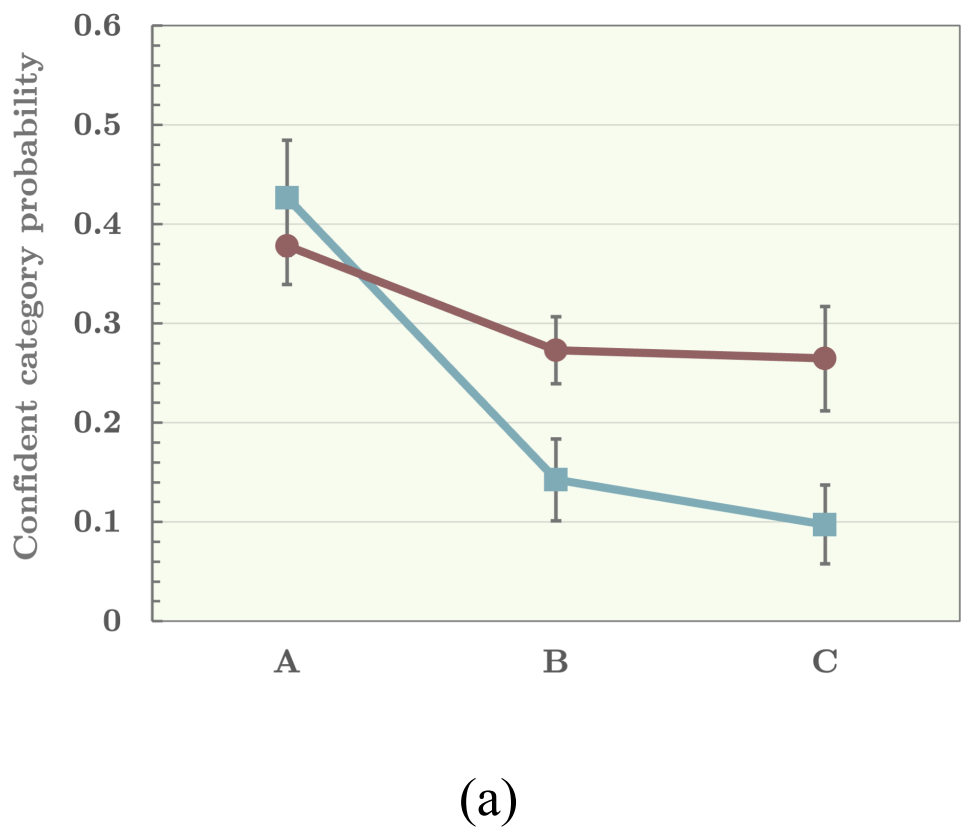

(a)

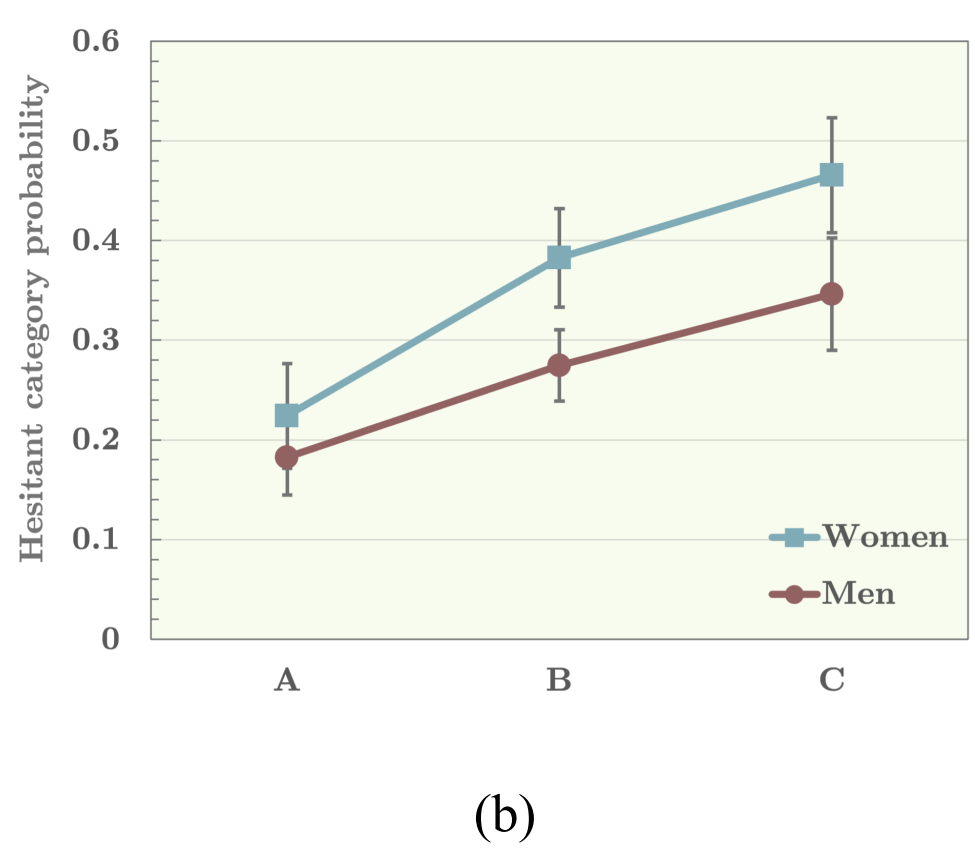

(b)

Figure 6: (a) Probability of being categorized as Confident at the beginning of Physics 2 for women and men with different Physics 1 grades. (b) Probability of being categorized as Hesitant at the beginning of Physics 2 for women and men with different Physics 1 grades.

The corresponding graph for the Hesitant category shows similar but more muted trends, and only the simple main effects of grade were statistically significant. In particular, lower grades in Physics 1 are associated with being more likely to be in the Hesitant category in Physics 2, but there was no significant differential reaction by gender, and the overall gender effect is small and not statistically significant. Thus, in contrast to ending in the Confident category, the large gender difference in transitions for the Hesitant category primarily reflects differences in grades received rather than differential reactions to the same grades.

## 5. General Discussion

### *5.1. RQ1. Does the distribution of students' mindsets about their own abilities in physics change over the course of taking an introductory physics course, and are there any gender differences in the distributions at each time point?*

Our study uses previously validated surveys to measure the extent to which students accept or reject a fixed mindset about their own physics abilities over a sequence of introductory physics courses to track changes for men and women separately. Examining the average survey scores on mindset items revealed an overall decline towards fixed beliefs for all students who advanced from Physics 1 to Physics 2, regardless of their gender. This finding is consistent with previous research on physics-specific mindset scores showing a decline over a semester of the introductory Physics 1 course (Malespina et al., 2022), and we observed a similar downward trend for students transitioning from Physics 1 to Physics 2.

Students in this course were provided with feedback on exams or weekly quizzes. Such feedback could have potentially supported the development of a growth mindset by helping students think more deeply about their mistakes and view them as part of the learning process. However, the direction of changes in fixed beliefs for our sample suggests otherwise. It is possible that students did not engage meaningfully with the available feedback, as no opportunities for submitting revised answers or re-taking the exam were provided. Additionally, the heavy emphasis on high-stakes exams, where many students continued to struggle, likely contributed to reinforcing fixed-mindset beliefs. Students may have interpreted poor performance or mistakes as a reflection of a lack of innate ability rather than viewing them as opportunities for reflection and growth. Such considerations are relevant to researchers across science disciplines, as they suggest that instructional features such as high-stakes assessment or limited opportunities for reflection and revision can systematically shape how students interpret feedback and, in turn, influence their beliefs about their potential in a field.

Studies in STEM education and particularly in physics have consistently reported negative shifts in students' psychological constructs following the completion of academically challenging courses. Such shifts have been repeatedly reported for students in math-intensive fields (Flanigan et al., 2017; Shively and Ryan, 2013; Limeri et al., 2020). This trend is not limited only to mindset, but also extends to students' selfefficacy, perceived recognition, and identity (Bottomley et al., 2022; Dou et al., 2016). The persistent decline in these correlated constructs highlights the importance of further reflection on current instructional practices and the effectiveness of implemented interventions. Our findings suggest that as students advance in the introductory physics courses, their responses show an increased fixed mindset view about their own abilities. Since even some students who received an A fell in the Hesitant category, we hypothesize that this could be associated with the specific classroom experience, such as interactions with other peers or the instructor. Alternatively, this decline could be related to students entering the course with unrealistic expectations or underestimating the required effort and time management skills, which could later lead to discouragement.

In terms of gender differences, while the declines in average mindset scores were similar for both women and men, women consistently scored lower than men with a medium effect size. This finding, aligned with previous research in STEM context, highlights that women are more likely than men to endorse a fixed mindset and associate success in physics with innate ability rather than hard work (Deiglmayr et al., 2019). The incoming differences in mindset also highlight that this gap has already been shaped through women's earlier experiences, which may include a lack of role models, societal stereotypes, or implicit biases leading to differences in prior preparation (Corbett and Hill, 2015; McGuire et al., 2020; Hazari et al., 2010). The difference in average mindset scores between men and women from the beginning of Physics 1 to the beginning of Physics 2 remained relatively stable, indicating no further increase in the pre-existing gender gap, but also no improvement in terms of closing it.

There were consistently about twice as many women in the Hesitant category, while only about a third as many women in the Confident category compared to men. Prior research in the same context has shown that moderately lower fixed mindset scores, such as those observed here in the Hesitant category, can still predict lower future performance, even after accounting for other academic predictors (Malespina et al., 2022). Thus, the distinction between these categories in our results, specifically the high and low ends, remains practically significant, given the disproportionate representation of women in the Hesitant and Confident groups. The significant gender differences in representation at the extremes of the distribution are similar, but with larger magnitudes, to the pattern seen in course grades. While our study is focused on introductory physics, this pattern may reflect broader disciplinary dynamics, especially within fields that emphasize innate talent and are known to exhibit greater under-representation of certain demographics (Leslie et al., 2015; Storage et al., 2016). In this sense, the gendered mindset distributions observed here may not be unique to physics but may arise in disciplines where success is framed in terms of innate ability.

### *5.2. RQ2. What patterns of physics mindset shifts occur, and how do these patterns differ between men and women?*

We examined transition patterns across categories for men and women separately to ensure that no changes go unaccounted for when individuals move in opposite directions between categories. The transition patterns showed that while nearly half of the women and men students remained in the same mindset category, the rest switched categories. We found significant gender differences in terms of staying within the Confident category or switching from Confident to Hesitant, both favoring men. This shows that women who begin as Confident in Physics 1 are less likely than men to stay within that category, and more than twice as likely as men to end up in the lowest mindset category, i.e., Hesitant.

Previous research in this context has highlighted how women tend to report more fixed mindsets regarding their physics abilities (Malespina et al., 2022). Our findings here provide a more nuanced view, showing that women experience a larger shift in their mindsets over time, with them being significantly less likely than men to remain confident in terms of rejecting fixed beliefs in physics. The trends indicate that the students, especially women, may not have received enough support to maintain a positive perception of their potential, or to embrace the idea that confusion and challenges are a normal part of the learning process. From a theoretical standpoint, our findings contribute to broader conversations in science education by providing empirical evidence for the contextual sensitivity of domain-specific mindsets, particularly in disciplines where stereotypes about ability or belonging are salient. For example, while our sample did not permit a sufficiently powered analysis by other demographic measures such as race/ethnicity, prior work has shown that certain groups facing stereotype threats, including African American students, are also under-represented in brilliance-focused fields (Storage et al., 2016). Therefore, our results support a more situated interpretation in which shifts in mindset may reflect students' evolving interpretations of their experiences within a disciplinary culture.

From a broader perspective, this also suggests that while many students' mindsets remained stable, the changing perceptions of their potential for success in physics over the course of a semester leave room for constructive interventions. In this context, unfortunately, these changes often reflect a shift toward more fixed beliefs. This pattern has implications beyond physics, as similar dynamics may emerge in other STEM disciplines with well-documented brilliance or demographic stereotypes (e.g., engineering and computer science) (Cheryan et al., 2017).

### *5.3. RQ3. Does grade predict different patterns of change in physics mindset, and do these relationships between grades and mindset change differ between men and women?*

Compared to men, the larger under-representation of women in the higher end of the mindset distribution, mirroring trends of women receiving fewer A grades and more C grades compared to men, raised questions about whether the likelihood of identifying as Confident or Hesitant differs for men and women by grade feedback with the same initial mindset. Logistic regression analysis revealed that overall, students became less likely to be categorized as Confident and more likely to be categorized as Hesitant, as their grades decreased from an A to a B or C. This trend was consistent across genders in terms of the direction of change in the likelihood. However, we observed a significant interaction between grade and gender variables in our model predicting the Confident category outcome. Controlling for the incoming mindset, receiving a grade of B or C appeared to reduce the probability of being categorized as Confident more sharply for women than for men.

Prior research has shown that women are more likely than men to drop out of STEM pathways even when passing the required courses; more specifically, women are more likely to switch out of STEM majors compared to men, following receipt of a C grade in a course (Seymour et al., 2019). Therefore, the differential decrease in probability of confidently rejecting fixed beliefs implies that similar grades may not be perceived similarly by men and women. In fact, women who get a B grade are even less probable than men with a C grade to end up in the Confident category.

Comparing our results with prior research, it is possible that these differences in changing mindsets may offer an explanation for why women tend to drop out of STEM pathways even when they have similar grades to men (Ellis et al., 2016). This suggests that high attrition rates for women compared to men may be driven more by diminished confidence in one's beliefs about their potential than actual differences in performance. Importantly, our findings are likely to extend beyond mindset to a broader set of constructs associated with persistence in STEM (Lv et al., 2022). For example, similar grade feedback may differentially shape sense of belonging, interest, and perceived value of the discipline for men and women, even when objective performance is comparable. Therefore, the observed grade by gender interaction may represent a more general mechanism through which disparities in persistence emerge.

We hypothesize that one possible explanation for this gendered reaction to grade feedback is that entrenched biases about physics and the lack of female role models in the field may lead women to rely more heavily on initial grades to decide if they have what it takes to excel in the discipline. Studies have shown that from an early age, women are generally less exposed to external validation of their physics potential, such as encouragement from family, instructors, or peers, and their decisions to pursue physics rely heavily on extrinsic motivation (Mujtaba and Reiss, 2013). This reliance has also been observed in engineering disciplines, where female engineers were more likely to associate success with external factors than non-engineering females, while this pattern was not observed for men (Heyman et al., 2002). Given the historical prevalence of male physicists, it may become more likely for women, even those initially interested in physics or considering it as a career, to turn to grades as a primary indicator of their abilities in the absence of external motivation. This reaction is especially concerning in introductory courses, where early challenges can overshadow academic potential and cause talented individuals to be prematurely discouraged.

### *5.4. Instructional Implications*

While our results highlight the importance of students' initial physics mindsets coming into the introductory course, the observed transition patterns and changes in overall mindset scores suggest the potential for shifts in students' attitudes. Our results indicate that the most common pathway in mindset transitions is staying within the initial category. This highlights the need to address existing issues earlier, both within educational systems and society, as efforts for reducing the gender gap in students' initial attitudes and beliefs towards learning have been shown to be effective in earlier stages of life (Blackwell et al., 2007; Porter et al., 2022; Law et al., 2021).

However, the pre-existing gap does not imply that any current issues should be overlooked simply because they have been longstanding or because earlier efforts were unsuccessful. Our findings show that even after controlling for prior mindset, women are still at a greater risk of endorsing a fixed mindset. This brings attention to the importance of interventions aimed at promoting a growth mindset in introductory physics courses, particularly in the early years of post-secondary education. Interventions proven to be effective for those with limited access to opportunities or low socioeconomic backgrounds (Sisk et al., 2018) can be used to foster a classroom environment that provides equal learning opportunities for all students and supports their individual growth.

Additionally, how students respond to receiving grades lower than what they are used to can play a key role in their decision to persist in their major (Seymour et al., 2019), particularly in introductory courses during the first years of college, in which students tend to receive lower grades compared to later academic years (Malespina et al., 2024b). Past research suggests that students' prior experiences with feedback during their pre-university education may influence how they respond to feedback in their first year of undergraduate studies (Robinson et al., 2013). This may lead women to react more negatively to poor performance in introductory courses in college, given their tendency to excel in school achievement (Voyer and Voyer, 2014). In addition to the discouraging impact of lower grades, interactions with instructors and faculty who endorse a fixed mindset can further hinder women's retention in STEM. This may risk turning courses such as introductory physics into "weed-out" courses, inadvertently overlooking untapped potential (Scherr et al., 2017; Fuesting et al., 2019). Therefore, engaging in productive feedback dialogues with students and exploring potential adjustments to their self-directed learning strategies can help students become more receptive to their grades and the idea of cultivating a growth mindset (Molloy et al., 2012; Carless, 2012).

In designing mindset interventions and feedback practices, it is necessary to consider how these elements are implemented and communicated with students. For example, rather than introducing a growth mindset as a direct guarantee of success, students should be informed that positive change is possible through effort, but it does not mean that it happens immediately or at the same pace for all students (Yeager and Dweck, 2020). Instructors and teaching assistants, who may also facilitate interventions, play a crucial role in communicating that information, especially when it comes to supporting students who are more susceptible to societal pressures and may need additional support and scaffolding along their academic journey (Zeeb et al., 2020; Canning et al., 2019). Furthermore, mindset interventions tend to be more effective when peer norms align with intervention messages, which further emphasizes the role of instructors in supporting productive student collaboration (Yeager et al., 2019).

Introductory physics courses would also benefit from being structured in a way that offers timely and constructive feedback from instructors, encouraging students to reflect on their progress across the semester. This can be achieved by having students analyze their performance on assessments and providing incentives to ensure meaningful engagement with course material (Webb and Paul, 2023). Having students reflect on their performance, alongside support from instructors who model and cultivate a supportive classroom culture aligned with growth mindset principles, may be effective in preventing declines towards fixed beliefs. Importantly, these instructional strategies are not limited to physics, as they are likely to be relevant across certain STEM and other disciplines where students may encounter early challenges and face brilliance or demographic stereotypes. Taken together, our findings position instructional design, especially feedback, assessment, and classroom norms, as a key mechanism that shapes student motivational trajectories in different educational contexts.

### *5.5. Limitations and Future Research*

In this paper, we explored the ways in which students' fixed physics mindsets progress over the course of taking an introductory calculus-based physics course, with a focus on potential gender differences in reaction to grades. However, we recognize that our results regarding changes in response to grades are correlational, even when including many important confounds (e.g., initial mindset), and therefore, the causal effects of grades are not fully established. In addition, relying solely on external variables such as grades cannot reveal the underlying mechanism of why certain patterns are observed. How students engage with feedback from instructors and the extent to which this engagement may influence gendered shifts in mindset patterns are among the additional factors that need further investigation. Conducting simultaneous interviews with a subset of students showing particular changes alongside quantitative analyses would also provide insights into underlying mechanisms.

Some limitations may also arise from the multimodal, skewed mindset data, as well as our categorization process. Because of the specific data distribution, the cutoff points used to define the mindset categories were close. In addition, restricting the sample to students with passing grades may limit generalizability for patterns that are not conditioned on grades, such as those addressed in our first and second research questions. However, we found consistent patterns when including all grade levels and incorporated pre-course to post-course mindset data. Therefore, this selectivity does not limit the generalizability of our findings regarding the effects of receiving various passing grades.

We also note that our results are based on a student sample studying introductory physics in a predominantly white, US-based institution. Therefore, our findings may not be generalizable to a different population in another field or institution. Future work could examine whether similar grade-related differences in various constructs emerge across other STEM disciplines and demographic groups, particularly in contexts where identity may shape how evaluative feedback is interpreted. Therefore, it would be useful to conduct a similar analysis in different contexts, such as in less selective or smaller universities, or for courses with different demographic compositions, to explore whether our results would translate.

## 6. Conclusion

With the recent focus on domain-specific mindsets and how they evolve in educational contexts (Limeri et al., 2020; Goldhorn et al., 2023; Spatz and Goldhorn, 2021; Limeri, 2025), our study brings to light the dynamic nature of students' fixed beliefs as they progress through introductory calculus-based physics courses, with a more fine-grained focus on the direction and magnitude of these changes between men and women and across grades. In this work, we found an overall decline in students' physics mindset towards fixed beliefs as they advanced from a calculus-based introductory Physics 1 to Physics 2 course, with large gender differences in representation at the low and high ends of the mindset categories, both favoring men. This trend mirrored the patterns observed in course grades, with more pronounced attitudinal differences compared to those of grades. Exploring patterns of transition across the mindset categories, we found that while the majority of students were likely to hold on to their initial mindset, women were significantly more likely to move away from the Confident category. Additionally, we found that as grades decrease, the shifts away from confidence in rejecting fixed beliefs tend to be larger in magnitude for women, despite involving a similar direction for all students. Our findings align with intelligence mindset theory, which posits that mindsets are shaped through experiences and interpretations of success and failure; without explicit support for reframing academic setbacks, students with lower grades may interpret them as evidence of fixed ability rather than as information for growth. Therefore, instructors must provide supportive learning environments in order to minimize these shifts towards fixed mindsets, and to recognize at-risk students in a timely manner to help retain individuals with potential for improvement in the field.

**Acknowledgement(s)**

Not applicable

**Competing interests**

The authors declare that they have no competing interests.

**Funding**

Not applicable

**Availability of data and materials**

The data presented in this study are available on request from the corresponding author due to the data privacy requirements of US FERPA regulations.